\documentclass{neus2025}
\usepackage{amsmath}
\usepackage{graphicx} 
\usepackage{color}
\usepackage{xspace}
\usepackage[most]{tcolorbox}
\usepackage{hyperref}
\usepackage{amssymb}
\usepackage{booktabs}
\usepackage{adjustbox}
\usepackage{paralist}
\usepackage{multirow}
\usepackage{booktabs}
\usepackage{todonotes}
\usepackage{wrapfig}
\usepackage{titlesec}

\tcbuselibrary{theorems}

\newtcbtheorem[
  auto counter,
  number within=section
]{promptbox}{Listing}{
  colback=gray!5,
  colframe=black!40,
  fonttitle=\bfseries
}{pr}

\newtcolorbox{graybox}{
  colback=lightgray,
  colframe=black,
  boxrule=0.5pt,
  arc=2pt,
  left=3pt,
  right=3pt,
  top=2pt,
  bottom=2pt,
  boxsep=2pt
}

\titlespacing*{\section}
{0pt}   
{8pt}   
{4pt}   

\titlespacing*{\subsection}
{0pt}
{6pt}
{3pt}

\titlespacing*{\subsubsection}
{0pt}
{6pt}
{3pt}

\newcommand{\tlguard}{\textsc{LTLGuard}\xspace}
\newcommand{\rafsl}{\textsc{RAFSL}\xspace} 

\newcommand{\always}{\textbf{G}\xspace}
\newcommand{\eventually}{\textbf{F}\xspace}
\newcommand{\until}{\textbf{U}\xspace}

\renewcommand{\next}{\textbf{X}\xspace}

\title[\tlguard: Formalizing LTL Specifications]{
\tlguard: Formalizing LTL Specifications with Compact Language Models and Lightweight Symbolic Reasoning
}
\usepackage{times}
\author{%
 \Name{Medina Andresel} \Email{medina.andresel@ait.ac.at}\\
 \Name{Cristinel Mateis} \Email{cristinel.mateis@ait.ac.at}\\
 \Name{Dejan Ni\v{c}kovi\'{c}} \Email{dejan.nickovic@ait.ac.at}\\
 \addr AIT Austrian Institute of Technology,  Vienna, Austria
 \AND
 \Name{Spyridon Kounoupidis} \Email{skoun@csd.auth.gr}\\
 \Name{Panagiotis Katsaros} \Email{katsaros@csd.auth.gr}\\
 \addr Aristotle University of Thessaloniki, Thessaloniki, Greece %
 \AND
 \Name{Stavros Tripakis} \Email{stavros@northeastern.edu}\\
 \addr Northeatern University, Boston, MA, USA%
}

\begin{document}

\maketitle
\thispagestyle{empty}

\begin{abstract}
Translating informal requirements into formal specifications is challenging due to the ambiguity and variability of natural language (NL). This challenge is particularly pronounced when relying on compact (small and medium) language models, which may lack robust knowledge of temporal logic and thus struggle to produce syntactically valid and consistent formal specifications. 
In this work, we focus on enabling resource-efficient open-weight models (4B--14B parameters) to generate correct linear temporal logic (LTL) specifications from informal requirements. We present \tlguard, a modular toolchain that combines constrained generation with formal consistency checking to generate conflict-free LTL specifications from informal input. Our method integrates the generative capabilities of model languages with lightweight automated reasoning tools to iteratively refine candidate specifications, understand the origin of the conflicts and thus help in eliminating inconsistencies. We demonstrate the usability and the effectiveness of our approach and perform quantitative evaluation of the resulting framework. 
\end{abstract}

\section{Introduction}
\label{sec:intro}

Formal specifications play an essential role in a wide range of verification-based activities. Yet, translating informal requirements to formal specifications remains a significant challenge and is widely recognized as one of the key barriers to the broader adoption of formal methods in industrial practice~\cite{10.1007/978-3-642-41010-9_5}. Mastering this requirements engineering activity requires specific skills and has a steep learning curve. Informal requirements are also by definition prone to multiple interpretations. Consider for example the specification formalism Linear Temporal Logic (LTL)~\cite{DBLP:conf/focs/Pnueli77} and the following informal requirement: \emph{every request is followed by a grant.} This under-specified requirement can have at least three different interpretations: every request is followed by a grant in (1) the same step, (2) the next step, or (3) a strict future. This example illustrates the lack of absolute truth when formalizing requirements -- each interpretation can be either correct or wrong depending on the implicit assumptions and the intended meaning of the informal requirement.  

Several approaches have emerged over the past few decades to facilitate the formalization of requirements~\cite{buzhinsky2019formalization}. The proposed methods include pattern-based specifications~\cite{dwyer1998property},~\cite{konrad2005automated},~\cite{Mavridou21}, boilerplates~\cite{c}, natural language parse trees~\cite{yan2015formal}, as well as statistical and neural machine translations~\cite{DBLP:conf/icse/HeBNIG22}. Despite this rich landscape of methods, the problem of translating informal requirements to formal specifications remains far from being fully resolved. 

The advent of Large Language Models (LLMs) has revolutionized software engineering. LLMs are well suited to help engineers translate informal requirements into formal specifications, as they can understand natural language context, detect domain-specific patterns, and produce structured outputs. Flagship models provide an impressive baseline for formalizing requirements. However, there are multiple concerns regarding their adoption in practice. First, these models are typically too large to be hosted locally, raising privacy concerns when sensitive requirements must be sent to externally hosted services. They are also energy-demanding due to their size, with questionable cost--benefit trade-offs. Finally, even flagship LLMs are prone to producing credible but semantically misleading hallucinations~\cite{DBLP:journals/corr/abs-2404-18930}. Proprietary LLMs provide limited control over decoding and internal safeguards, making it difficult to reliably constrain outputs. In contrast, compact open-weight models can be hosted on local machines, but they often exhibit lower reliability on niche, logic-heavy tasks such as requirements formalization, including a higher propensity for unsupported or inconsistent outputs~\citep{ul-islam-etal-2025-much}.

We present \tlguard, a privacy-preserving framework for translating informal requirements into LTL using compact language models. These models are not necessarily trained on temporal logic and do not match the out-of-the-box translation accuracy of large flagship models. Rather than relying on computationally intensive methods such as fine-tuning, we show how lightweight techniques, including syntax-constrained decoding, retrieval-augmented few-shot learning and consistency checking, can substantially improve the robustness and translation accuracy of these compact models. In doing so, we show that reliable and mutually consistent formalizations~(\cite{mokos26}) can now be achieved through a principled integration of lightweight LLM prompts and automated reasoning, without specialized training or large-scale model adaptation.

\section{Related Work}
\label{sec:related-work}

In addition to the works mentioned in Section~\ref{sec:intro}, there is a plethora of recent LLM-based methods for translating informal requirements to formal specifications. 
\cite{DBLP:conf/aaai/FuggittiC23} present NL2LTL for translating NL into LTL through mapping inputs to predefined patterns, such as DECLARE templates, and using filtering functions to rank candidate formulas based on conflicts or subsumptions. We leverage compact language models for more flexible specification generation and an automated reasoning–based refinement loop to help users resolve semantic inconsistencies that fall outside a fixed set of patterns.
\cite{DBLP:conf/cav/CoslerHMST23} introduce nl2spec, an \emph{interactive} LLM-based framework that translates NL requirements into temporal logics (including LTL), with a strong emphasis on human-in-the-loop disambiguation. 
Their key idea is to decompose a requirement into fine-grained subformulas, enabling iterative disambiguation while checking syntactic validity through downstream parsing. In contrast, our approach emphasizes syntax-constrained generation and automated consistency checking across sets of specifications to resolve cross-formula conflicts rather than  per-formula semantic debugging.
\cite{kogler2024reliable} improve the reliability of LLM-generated specifications by enforcing strict syntactic constraints during token generation through an automaton-based streaming post-processor. 
Their framework employs a deterministic finite automaton (DFA) to validate candidates in real-time against a JSON schema, ensuring the output adheres to a formal Domain-Specific Language (DSL).
Rather than limiting control to token-level syntactic filtering, we combine automatic syntax correction with an iterative refinement loop and automated reasoning to surface semantic issues and logical conflicts in translated temporal specifications.
\cite{DBLP:conf/qrs/XuFM24} introduce an interactive NL-to-LTL translation approach that uses embedding-based retrieval and dynamic prompt evolution, incrementally enriching few-shot examples with user-corrected translations to improve per-requirement accuracy without fine-tuning. Complementarily, our approach also supports modular example retrieval and incremental dataset enrichment, but places stronger emphasis on syntax-constrained generation and automated reasoning to ensure consistency across sets of LTL specifications rather than focusing solely on individual translations.
\cite{ma2025bridging} propose REQ2LTL, a two-stage framework mitigating end-to-end generation errors by decomposing requirements into a hierarchical intermediate representation via an LLM and then translating them into syntactically correct LTL. We operate in a compact-model setting targeting cross-specification coherence rather than only per-requirement structural fidelity. We
combine syntax-constrained decoding and iterative repair with automated reasoning for consistency checking LTL formulas and guiding human conflict resolution.
Ma et al. report results with large LLM backends but do not release the full implementation and industrial dataset, limiting direct comparison in our compact language models setting.
\cite{DBLP:conf/rv/CohenHPG25} propose LLMon, an interactive framework for synthesizing LTL Past monitors from NL requirements, generating multiple semantic interpretations (including executable monitors) and using distinguishing traces to help users resolve ambiguity at the level of individual operators. While LLMon leverages solely LLMs in every step of the translation process, our approach emphasizes the combination of compact models with symbolic methods across requirements.
\cite{DBLP:conf/iccps/ShuklaTP25} propose a gray-box NL-to-TL translation approach that conditions LLMs on structured intermediate representations, combining curated examples and fine-tuning to improve single-formula translation accuracy, with correctness assessed via downstream verification tools. In contrast, our work avoids fine-tuning and instead emphasizes more lightweight methods for increasing the translation accuracy.

\section{Framework for Formalizing Informal Requirements}
\label{sec:methodology}

This section introduces our framework for formalizing requirements into LTL specifications. In Section~\ref{sec:problem} we state the problem that we solve and the desired properties of an appropriate solution. Section~\ref{sec:overview} describes an overview of our framework, while Section~\ref{sec:blocks} details its building blocks.

\subsection{Problem Definition and Solution Requirements}
\label{sec:problem}

The requirements formalization problem to be addressed is the following: \emph{Given a set of functional requirements expressed in natural language, the goal is to derive a corresponding set of LTL specifications that faithfully capture the intended semantics and are mutually consistent.}

We seek a solution based on a compact language model that can be used on a local computing infrastructure and that does not require specialized expertise or large-scale model tuning. We identify \emph{correctness} and \emph{robustness} as the two most important properties for such a solution.

\paragraph{Correctness.}  We consider three dimensions of correctness. The first dimension is \emph{syntactic correctness}: the generated output has to be a syntactically valid LTL formula. The second dimension is the \emph{semantic correctness}: the formula should preserve the intended semantics of the informal requirement. While semantic correctness is fundamental, it cannot be fully evaluated in practice due to the absence of an authoritative ground truth for informal-to-formal translations. The third dimension is the \emph{ambiguity handling}: the solution shall resolve ambiguous formulations of NL requirements that may admit multiple semantically valid formalizations, depending on their interpretation.

\paragraph{Robustness.} Robustness refers to the stability of the translation with respect to \textit{close variations} of the input. 
We identify two notions of robustness, explained in Section~\ref{sec_robustness_eval}.

\subsection{Overview of the Framework}
\label{sec:overview}

\begin{wrapfigure}{l}{0.75\linewidth}
\centering
\includegraphics[width=\linewidth]{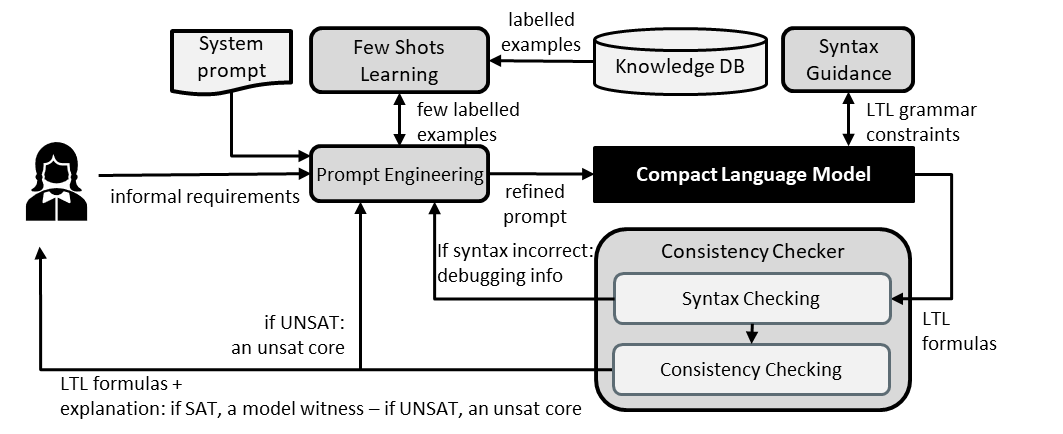}
\caption{Overview of the specification formalization approach.}
\label{fig:overview}
\end{wrapfigure}

Figure~\ref{fig:overview} depicts the proposed methodology.
The toolchain provides semi-automated support for formalizing requirements while keeping the human in the loop. This choice is motivated by the absence of a definite ground truth for correct formalizations. The workflow starts with a prompt written by the engineer, requesting to translate a set of informal requirements into LTL specifications. A separate system prompt supplies the necessary background context for the task. The user prompt is further enriched with additional NL-LTL pairs retrieved by a retrieval-augmented few-shot learning (\rafsl) module. \rafsl dynamically selects relevant examples from a dataset and incorporates them into the prompt, thereby providing task-specific context to the translator.

\noindent The enriched prompt is then provided to a compact language model. Its output is constrained by a syntax-guidance module that leverages the LTL grammar to steer the model toward syntactically correct translations. Generated candidate formulas are subsequently checked by an LTL parser to ensure syntactic validity. If parsing fails, the debugging information is fed back to the model to support iterative refinement toward a syntactically correct formula. Once formulas pass the syntax checks, a consistency checker verifies that they do not contain mutual conflicts, which may indicate either genuinely inconsistent requirements or inaccuracies in their formalization. To help the engineer understand the source of the issue, the toolchain highlights conflicting requirements and provides the corresponding unsatisfiable core computed by the consistency checker. For inconsistency cases, this diagnostic information is fed back to the model in an attempt to automatically resolve the conflicts.

\subsection{Building Blocks of the Framework}
\label{sec:blocks}

\subsubsection{Prompt Specification}
\label{sec:prompt}

The first step in ensuring correct and robust translation of NL requirements using a language model is the design of a \emph{prompt specification}. It defines a system prompt that constrains and guides an LLM’s behavior by explicitly stating its role, assumptions, and expected output structure in order to achieve a desired task. Prompt specification is typically an iterative activity, which requires multiple refinements of the system prompt that is appropriate for the specific task. 
We converged to the system prompt shown in Listing~\ref{pr:system}. It defines the role of the model, provides syntactic guidelines about LTL and scopes the formatting of the model output. 

\begin{promptbox}{System Prompt}{system}
\footnotesize
You are an expert in formal verification and temporal logic. You will be given textual requirements. Your task is to translate each requirement into LTL formula(s) using the following conventions:

- Temporal operators: G = always, F = eventually, X = next, U = until.

- Atomic propositions must be lowercase words (e.g., request, granted), not single uppercase letters.

- Output strictly and only the LTL formula(s). Do not include reasoning, explanations, steps, or natural language of any kind. The output should contain only the formula(s), nothing else.

- If the provided text is not related to LTL requirements at all, just output "The provided text has nothing to do with LTL".
\end{promptbox}

\subsubsection{Retrieval-Augmented Few-Shot Learning (\rafsl)}
\label{sec:fsl}

Smaller LLMs often have higher hallucination rates and lower accuracy in niche tasks compared to large proprietary LLMs~\citep{ul-islam-etal-2025-much}. When used for requirements formalization, smaller LLMs may produce semantically or syntactically incorrect formulas, due to their limited exposure to temporal logic during pre-training. To compensate for these shortcomings, we employ a retrieval-augmented few-shot learning (\rafsl) approach, which dynamically provides the model with relevant examples at inference time.

In contrast to common few-shot learning techniques that add a predefined fixed set of examples directly to the prompt~\citep{DBLP:journals/corr/abs-2005-14165,dong-etal-2024-survey}, our approach leverages dense retrieval to dynamically select examples based on semantic similarity for each input requirement. Given a NL requirement as input, we encode it into an embedding using a sentence transformer. We then use this embedding to search a pre-built index of NL-LTL pairs, where each NL requirement has been encoded similarly to the input requirement. We retrieve the top-k most similar examples based on the highest cosine similarity to the query and then add these examples to the prompt, providing the LLM with relevant examples before it generates the LTL formula.

To construct our retrieval dataset of NL-LTL pairs, we selected some of 
the temporal property patterns from \cite{dwyer1998property}, which provide a catalog of common temporal property patterns along with their LTL formalizations. We selected only patterns where their corresponding LTL formulas were deemed concise, aiming for easily interpretable LTL formulas. We then expanded the dataset with NL-LTL pairs drawn from the formal verification literature.
We abstracted all atomic propositions in both the NL requirements and their corresponding LTL formulas to generic placeholders (atom\_1, atom\_2, etc.), enabling the retrieval mechanism to match based on temporal structure rather than specific variable names. To address the inherent ambiguity of NL and improve robustness, we introduced paraphrasing into the dataset, ensuring that our approach retrieves relevant examples regardless of the exact requirement formulation.

\subsubsection{Grammar-based Guidance}
\label{sec:guidance}

One big challenge in employing LLMs to formalize text-based requirements into correct LTL formulas is ensuring the output is syntactically valid w.r.t. LTL input grammar. In order to achieve this, we employ the following strategies: \begin{inparaenum}[(i)]
\item adding the grammar in the system prompt and provide clear instructions,
\item applying grammar-based decoding tools, and
\item combination of (i) and (ii).
\end{inparaenum}

We instantiate our framework with the grammar-based decoding tool SynCode (\cite{DBLP:journals/tmlr/UgareSKM025}), a state-of-the-art approach that uses a deterministic finite automaton (DFA) lookup, created based on a given input grammar, to apply a masking decoding approach for enabling the LLM in choosing the next token such that it is complying with the input grammar. 
This is done by pre-computing a DFA mask store, a lookup table derived from the grammar’s DFA that maps grammar states to valid tokens. This enables SynCode to quickly filter invalid tokens, enabling efficient grammar-constrained decoding. 
While the approach is sound and complete under certain conditions, a complete, correct result is not always guaranteed.

\subsubsection{Consistency Checking with Syntax Checking and Explanations}
\label{sec:consistency}

A consistency checker serves to determine whether an LTL formula is \emph{satisfiable}, i.e., whether there exists at least one execution trace that satisfies all its temporal constraints. The checker typically returns one of two outcomes: SAT, meaning the formula is consistent, together with a model of the formula; or UNSAT, meaning the formula is inconsistent, together with an unsatisfiable core. A model that witnesses the satisfiability of an LTL formula is an infinite trace for which the formula holds along the entire sequence. An unsatisfiable core is a minimal (or smaller) subset of the original formula that is itself inconsistent, thereby explaining why the full formula cannot be satisfied. The model and the unsatisfiable core both provide valuable information that explains the outcome of the analysis to the human user. This information can be additionally used as feedback for improving the translation process. We use BLACK~\cite{DBLP:conf/gandalf/GeattiGM21} as our satisfiability checker for LTL.

\section{Evaluation and Experimental Results}
\label{sec:eval}

\subsection{Research Questions}

\paragraph{(RQ1) Translation Accuracy.} What is the accuracy of \tlguard in translating natural language requirements to LTL and how much each component contributes to it?

\noindent\textit{Rationale}:
This research question addresses the ability of \tlguard to bridge the gap between informal requirements and their associated formal specifications. It also quantifies the individual impact of each module in \tlguard on the overall accuracy of the translated LTL specifications.
\noindent\textit{Metrics}:
We evaluate the accuracy of \tlguard by measuring its syntactic and semantic accuracy on benchmarks consisting of labeled NL requirement - LTL formulas. We measure the contribution of each module using an ablation study, i.e. by systematically disabling each component and comparing the resulting change in specification accuracy against the full system baseline.

\paragraph{(RQ2) Robustness.} How robust is \tlguard to linguistic variation in NL requirements?

\noindent\textit{Rationale}: 
Evaluating robustness provides an answer whether \tlguard maintains stable performance when NL requirements are subject to linguistic variability.
\noindent\textit{Metrics}: 
Robustness is measured by comparing the accuracy of the generated LTL specifications across multiple paraphrased variants of the same requirements.

\paragraph{(RQ3) Consistency Checking.} To what extent can \tlguard detect logical inconsistencies in natural language requirements?

\noindent\textit{Rationale}: 
This research question investigates the effectiveness of \tlguard in identifying logical conflicts that arise either from inherently inconsistent requirements or from erroneous formalizations.
\noindent\textit{Metrics}: We measure the ability of \tlguard to detect conflicts over a small benchmark that contains both consistent and intentionally inconsistent sets of requirements.

\subsection{Experimental Setup}

\paragraph{Implementation.}
\tlguard is implemented in Python 3.10.18 using SynCode 0.4.16, LangChain  1.0.7, Faiss-cpu 1.13.1 and the latest version\footnote{As of February 2026.} of LTL solver BLACK. To facilitate performing the ablation study and providing an answer to (RQ2), we implement 7 variants of our framework. Table~\ref{tab:ablation-all} shows for each variant the combination of enabled and disabled components. All experiments were run on a high-performance computing cluster with 3 nodes with 64 CPUs (Intel(R) Xeon(R) Silver, 16 cores per CPU) and  6 GPUs (2 NVIDIA TITAN RTX, 4 NVIDIA GeForce RTX 2080).

\paragraph{Datasets.}
We use three datasets in our experiments:
(i) The ablation study (RQ1) uses a curated set of 70 NL-LTL pairs representing common requirements and their linguistic variations: it covers predominantly safety and response-style specifications, includes bounded-response with $\next$-nesting, mutual-exclusion constraints, conjunctions of multiple properties, and a small number of additional operators (e.g., $\until$); multiple distinct NL paraphrases map to the same underlying formula, enabling us to probe sensitivity to phrasing and proposition naming.
(ii) For comparison to prior work, we use the 36-instance ``hard'' benchmark from \texttt{nl2spec}~\cite{DBLP:conf/cav/CoslerHMST23}.
(iii) Finally, \rafsl relies on a knowledge base of lifted NL-LTL example pairs (Section~\ref{sec:fsl}), used in two versions: a full set (137 pairs) and an overlap-removed set (122 pairs) to quantify the effect of example overlap with the \texttt{nl2spec} benchmark.
The robustness (RQ2) and consistency-checking (RQ3) evaluations use only the small, explicitly listed requirement sets introduced in their respective sections.

\subsection{Evaluation Results}

\subsubsection{(RQ1) Translation Accuracy}

\paragraph{Translation accuracy and ablation study.} The  first experiment assesses the effectiveness of our framework in improving (i) syntactic validity of generated LTL formulas and (ii) semantic correctness with respect to an expected specification.
We perform an ablation study across four compact open-weight LLMs available on HuggingFace:
Qwen2.5-14B-Instruct~\cite{hf:qwen2.5-14b},
Mistral-Nemo-Instruct-12B~\cite{hf:mistral-nemo},
Mistral-7B-Instruct~\cite{hf:mistral-7b}, and
Phi-3-Mini-4B-Instruct~\cite{hf:phi3-mini}.
To generate sentence embeddings used in the \rafsl component, we employ Qwen3-Embedding-0.6B~\cite{hf:qwen3-embedding-0.6b}. 
For each model, we evaluate seven variants that progressively enable components of our pipeline. Table~\ref{tab:ablation-all} summarizes the results.

\begin{table}[tbh]
\centering
\footnotesize
\setlength{\tabcolsep}{3.2pt}
\renewcommand{\arraystretch}{0.95}
\begin{adjustbox}{max width=\linewidth}
\begin{tabular}{l|cccc|cc|cc|cc|cc}
\toprule
\multirow{2}{*}{\textbf{\tlguard Variant}} &
\multicolumn{4}{c|}{\textbf{Components}} &
\multicolumn{2}{c|}{\textbf{Qwen2.5-14B}} &
\multicolumn{2}{c|}{\textbf{Mistral-Nemo-12B}} &
\multicolumn{2}{c|}{\textbf{Mistral-7B}} &
\multicolumn{2}{c}{\textbf{Phi3-mini-4B}} \\
& \textbf{G} & \textbf{S} & \textbf{R} & \textbf{F} &
\textbf{Syn.} & \textbf{Sem.} &
\textbf{Syn.} & \textbf{Sem.} &
\textbf{Syn.} & \textbf{Sem.} &
\textbf{Syn.} & \textbf{Sem.} \\
\midrule
V1: Vanilla (LLM only + G-free Prompt)           & $\times$ & $\times$ & $\times$ & $\times$ & 95.7 & 68.5 & 51.4 & 31.4 & 10.0 &  7.1 & 47.1 & 24.2 \\
V2: Prompt Grammar               & \checkmark & $\times$ & $\times$ & $\times$ & 57.1 & 34.2 & 17.1 &  5.7 &  5.7 &  1.4 & 35.7 & 11.4 \\
V3: Grammar + SynCode            & \checkmark & \checkmark & $\times$ & $\times$ & 97.1 & 50.0 & 85.7 & 27.1 & 15.7 &  5.7 & 55.7 & 15.7 \\
V4: Grammar + SynCode + \rafsl   & \checkmark & \checkmark & \checkmark & $\times$ &
95.7 & 75.7 & \textbf{92.8} & \textbf{67.1} & 87.1 & 38.5 & 90.0 & 60.0 \\
V5: Grammar + SynCode + Feedback & \checkmark & \checkmark & $\times$ & \checkmark &
\textbf{98.5} & 52.8 & 85.7 & 28.5 & 42.8 &  7.1 & 65.7 & 13.0 \\
V6: SynCode + \rafsl + Feedback  & $\times$ & \checkmark & \checkmark & \checkmark &
97.1 & \textbf{78.6} & \textbf{92.8} & 64.2 & 78.5 & \textbf{40.0} & 91.4 & \textbf{64.2} \\
V7: Full System                  & \checkmark & \checkmark & \checkmark & \checkmark &
\textbf{98.5} & 74.2 & \textbf{92.8} & 57.1 & \textbf{92.8} & 38.5 & \textbf{92.8} & 35.7 \\
\bottomrule
\end{tabular}
\end{adjustbox}
\caption{Framework variants and ablation results. \textbf{G}: prompt grammar, \textbf{S}: SynCode strict decoding, \textbf{R}: \rafsl, \textbf{F}: parser feedback (\checkmark enabled, $\times$ disabled). \textbf{Syn.}=\% syntactically valid LTL; \textbf{Sem.}=\% semantically correct w.r.t.\ the labels (accepting logically equivalent formulas).}
\label{tab:ablation-all}
\end{table}

We observe that enabling lightweight components of our pipeline (no fine-tuning and no full-blown RAG, and using only a limited number of examples) substantially improves output quality for all evaluated models.
In particular, the gains are largest for models with weaker prior knowledge of LTL.
For Mistral-7B, syntactic validity increases from 10.0\% (V1) to 92.8\% (V7), and semantic correctness (equivalence-aware) increases from 7.1\% (V1) to 40.0\% (V6) / 38.5\% (V7).
Similarly, Phi3-mini-4B improves from 47.1\% to 92.8\% syntactic validity, and from 24.2\% to 64.2\% semantic correctness when enabling V6 (with the best semantic score for this model).
For Mistral-Nemo-12B, the best-performing configuration reaches 92.8\% syntactic validity and 67.1\% semantic correctness (V4), compared to 51.4\% and 31.4\% in the vanilla setting.
For Qwen2.5-14B, which already performs strongly in the vanilla setting, the pipeline still yields consistent improvements in semantic correctness (68.5\% $\rightarrow$ 78.6\% in V6) while maintaining near-perfect syntactic validity.
Across models, variants that combine SynCode with retrieval of relevant examples (V4, V6, V7) tend to yield the strongest semantic improvements, suggesting that supplying targeted context is especially beneficial when the base model has limited latent knowledge of temporal logic.

\begin{table}[tbh]
\centering
\small

\footnotesize
\setlength{\tabcolsep}{4pt}
\renewcommand{\arraystretch}{0.95}
\begin{tabular}{l|l|l}
\hline\textbf{Sentence} & \textbf{Expected} & \textbf{Actual} \\
\hline
The system shall eventually reach a safe state. & $\always\,\eventually\,safe$ & $\eventually\,safe$ \\
$p$ and $q$ should not be both true. & $\always\,\neg\,(p \land q)$ & $\neg\,(p \land q)$ \\
After $p$, always $q$. & $\always\,(p \to \always\,q)$ & $\always\,(p \to \next\,\always\,q)$ \\
\hline
\end{tabular}
\caption{Ambiguous requirements where the generated formula differs from the expected one.}
\label{tab:ambiguity-examples}
\end{table}

Semantic correctness in Table~\ref{tab:ablation-all} was computed using an equivalence-aware criterion, where generated formulas were counted as correct when logically equivalent to the expected formula.
Nevertheless, NL ambiguity remains a practical challenge: a requirement may allow multiple non-equivalent yet plausible formalizations under different interpretations, as illustrated in Table~\ref{tab:ambiguity-examples} with examples from our experiment.

\paragraph{\texttt{nl2spec} ``hard'' benchmark.}

We evaluated our best-performing setup (V6 with \textsc{Qwen2.5-14B}) on the ``hard'' benchmark from~\cite{DBLP:conf/cav/CoslerHMST23}, who
 report semantic accuracy for multiple \texttt{nl2spec} setups (Section~4.2, Table~1), including initial (non-interactive) translations under different prompting strategies and an interactive setting with user-guided corrections.
Their evaluation counts an instance as correct only if it matches the expert's intended meaning, i.e., it is ambiguity-intolerant in the sense that alternative plausible interpretations of ambiguous NL input are not accepted.

\noindent\textit{Dataset irregularity.} 
We identified one requirement in the hard dataset that is not comprehensible English (“a must always hold, but if is exceeds, it allow two timestamps to recover”) and whose meaning could not be inferred even with the reference formula; consequently, no configuration of \tlguard reproduced the expected label, and we report results on the full dataset while noting that this instance likely constitutes an unavoidable error for any translation method.

\noindent\textit{Role of \rafsl overlap and evaluation strategy.}
We observed overlap between our \rafsl dataset and the \texttt{nl2spec} benchmark: 15/137 examples (11\%) match 15/36 hard instances (42\%) in template structure (modulo atom name). Such overlap is expected due to standard requirements formulations and recurring patterns. To assess its impact, we ran two experiments: Exp.~1 (overlap) uses the full \rafsl dataset, while Exp.~2 (no overlap) removes overlapping examples before retrieval. 
Semantic correctness is evaluated under two strategies: S1 (ambiguity-intolerant) counts only predictions matching the benchmark’s single expert interpretation~\cite{DBLP:conf/cav/CoslerHMST23}, whereas S2 (ambiguity-friendly) also accepts plausible alternative formalizations. Table~\ref{tab:nl2spec-compare-qwen} reports accuracies and compares \tlguard to the systems reported by Cosler et al.\ on the same benchmark.

\begin{table}[tbh]
\centering
\small
\setlength{\tabcolsep}{4pt}

\begin{adjustbox}{max width=\linewidth}
\footnotesize
\setlength{\tabcolsep}{4pt}
\renewcommand{\arraystretch}{0.95}
\begin{tabular}{l|c|c|cc}
\toprule
\textbf{Setup (36 hard instances)} & \textbf{Backend} & \textbf{Syntax (\%)} & \multicolumn{2}{c}{\textbf{Sem. (\%)}} \\
& & & \textbf{S1} & \textbf{S2} \\
\midrule
\texttt{NL2LTL}~\cite{DBLP:conf/aaai/FuggittiC23}                          & Rasa     & --   & 2.7  & -- \\
T5 fine-tuned~\cite{DBLP:journals/corr/abs-2206-01962}                     & T5       & --   & 5.5  & -- \\
\midrule
\texttt{nl2spec} initial~\cite{DBLP:conf/cav/CoslerHMST23}                 & Bloom    & --   & 13.8 & -- \\
\texttt{nl2spec} initial~\cite{DBLP:conf/cav/CoslerHMST23}                 & Codex    & --   & 44.4 & -- \\
\texttt{nl2spec} initial + ID examples~\cite{DBLP:conf/cav/CoslerHMST23}   & Codex    & --   & 58.3 & -- \\
\texttt{nl2spec} interactive~\cite{DBLP:conf/cav/CoslerHMST23}             & Codex    & --   & 86.1 & -- \\
\texttt{nl2spec} initial~\cite{DBLP:conf/cav/CoslerHMST23}                 & GPT-3.5  & --   & 33.3 & -- \\
\texttt{nl2spec} initial + ID examples~\cite{DBLP:conf/cav/CoslerHMST23}   & GPT-3.5  & --   & 47.2 & -- \\
\texttt{nl2spec} interactive~\cite{DBLP:conf/cav/CoslerHMST23}             & GPT-3.5  & --   & 58.3 & -- \\
\midrule
\textbf{\tlguard{:} V6 + \rafsl overlap (Exp.~1)}                           & Qwen2.5-14B & \textbf{100.0} & \textbf{75.0} & \textbf{77.8} \\
\textbf{\tlguard{:} V6 + \rafsl no overlap (Exp.~2)}                        & Qwen2.5-14B & \textbf{97.2} & \textbf{50.0} & \textbf{63.9} \\
\bottomrule
\end{tabular}
\end{adjustbox}
\caption{Results on the 36-instance \texttt{nl2spec} ``hard'' benchmark. \textbf{Syntax}=\% syntactically valid LTL; \textbf{Sem.}=\% semantic accuracy under \textbf{S1} (ambiguity-intolerant; expert-aligned) and \textbf{S2} (ambiguity-friendly).
For \texttt{nl2spec}, \texttt{NL2LTL}, and T5 rows we reproduce Cosler et al.'s Table~1 S1 accuracies; Syntax and S2 are not reported there and are shown as ``--''.}
\label{tab:nl2spec-compare-qwen}
\end{table}

\noindent
\textit{Interpretation.} Several trends emerge.
First, \tlguard achieves strong performance on this challenging benchmark with a compact model, whereas the highest accuracies reported for \texttt{nl2spec} in~\cite{DBLP:conf/cav/CoslerHMST23} are obtained with substantially larger proprietary LLMs (Codex and GPT-3.5) and, in the best case, with interactive user debugging.
In particular, with \rafsl overlap (Exp.~1), \tlguard reaches at least 75.0\% semantic accuracy under both the ambiguity-intolerant (S1) and ambiguity-friendly strategy (S2), while maintaining perfect syntactic validity, outperforming all non-interactive \texttt{nl2spec} variants and approaching the accuracy of the best-performing interactive \texttt{nl2spec} with Codex (86.1\%).
This suggests that careful prompt structuring, syntax guarantees, and RAFSP can substantially close the gap to larger LLMs, even without fine-tuning.

Second, comparison to other baselines further highlights \tlguard effectiveness.
\texttt{NL2LTL} (pattern-driven translation)~\cite{DBLP:conf/aaai/FuggittiC23} achieves 2.7\% accuracy on the same hard set, and the fine-tuned T5 approach~\cite{DBLP:journals/corr/abs-2206-01962} (T5-base, 220M parameters) reaches 5.5\%.
\texttt{nl2spec} with an open-weight but very large model (Bloom-176B) attains only 13.8\%.
\tlguard\ -- using a substantially smaller open-weight model (Qwen2.5-14B) -- achieves semantic accuracy from 50.0\% to 77.8\% depending on overlap and ambiguity handling, indicating that our toolchain can make moderate-size models competitive on challenging NL-to-LTL translation tasks.

Third, removing overlap (Exp.~2) causes S1 semantic accuracy to drop to 50.0\%, while syntactic accuracy remains high (97.2\%). This suggests the hard benchmark includes patterns where \rafsl examples are crucial, i.e. prompting and constrained decoding alone cannot reliably recover the expert’s intended interpretation.  
The S1–S2 gap in Exp.~2 (50.0\% vs.\ 63.9\%) indicates that many ``errors'' under the benchmark's ambiguity-intolerant scoring reflect plausible alternative interpretations rather than true translation failures. Thus, strict expert-alignment metrics may underestimate practical usefulness without explicit ambiguity handling (e.g., via multiple candidate formalizations or gold formulas).

\noindent
\textit{Illustrative effect of \rafsl wrt ambiguity.} 
For ``Whenever $a$ and $b$ do not hold, $c$ holds eventually.'', the benchmark label is 
\(\always(\neg\,(a \land b) \to \eventually\,c)\).
We consider the sentence ambiguous: the plural form ``do not hold'' may mean \(\neg\,a \land \neg\,b\), whereas in practice it is sometimes interpreted as \(\neg\,(a \land b)\).
With overlap, \rafsl retrieves a lifted near-duplicate pair
\(\langle\)``Whenever atom\_1 and atom\_2 do not hold, atom\_3 holds eventually.'', \(\always(\neg\,(atom\_1 \land atom\_2) \to \eventually\,atom\_3)\)\(\rangle\), 
and \tlguard matches the label; without overlap it instead yields
\(\always((\neg\,a \land \neg\,b) \to \eventually\,c)\),
aligning with the alternative interpretation.
Moreover, changing the stored label for the retrieved lifted example systematically steers generation (e.g., to \(\always(\neg(a \land b) \to \next\,\eventually\,c)\) for a ``strictly future'' reading, or to \(\always(\neg(a \land b) \to (\neg(a \land b)\ \until\,c))\) to require $c$ before $(a\land b)$ becomes true again), illustrating how \rafsl can encode and propagate domain-specific interpretation choices.

\begin{graybox}
\textbf{(RQ1):} \tlguard couples strong syntactic guarantees with a modular retrieval mechanism that improves semantic accuracy on difficult benchmarks. Overlap with stored examples significantly affects outcomes. Ambiguity-aware generation is crucial for realistic requirements formalization.
\end{graybox}

\subsubsection{(RQ2) Robustness Evaluation}
\label{sec_robustness_eval}

We evaluate the robustness of our NL-to-LTL translation using two notions of robustness: (1) {\em Atom renaming robustness} (ARR) and (2) {\em Rephrasing robustness} (RR). The idea behind ARR is that when the structure of the English phrase remains the same, and only the words mapped to atomic propositions change, the resulting LTL formulas should be identical modulo renaming of the atomic propositions. For example, all requirements in Table~\ref{tab:robust_sets}, under column ARR, should be translated to the same LTL formula template.  Ideally, they should all be translated to the correct request-response property $\always(p\to\eventually q)$, where $p$ and $q$ change for each requirement. 
But note that robustness is orthogonal to correctness. We would therefore still qualify as ARR robust (albeit incorrect) a translation which translates all these requirements into LTL formulas of the form, say, $\always\eventually(p\land q)$.

\begin{table}[tbh]
\centering
\footnotesize
\setlength{\tabcolsep}{4pt}
\renewcommand{\arraystretch}{1.0}
\begin{tabular}{p{0.38\linewidth} || c | p{0.42\linewidth}}
\toprule
\textbf{ARR} & \textbf{Group} & \textbf{RR} \\
\midrule
1. Every request is eventually granted. 
& \multirow{3}{*}{G1} 
& a. Every request is eventually granted. \\

2. Every message is eventually delivered. 
& 
& b. No request remains ungranted. \\

3. Every button press is eventually processed. 
& 
& c. Every request will eventually be granted. \\

\cmidrule(lr){2-3}

4. Every order is eventually shipped. 
& \multirow{2}{*}{G2} 
& d. Every message is eventually delivered. \\

5. Every error is eventually logged. 
& 
& e. No message is left undelivered. \\

6. Every task is eventually completed.
& 
& f. Every message will sooner or later be delivered.

\\
\bottomrule
\end{tabular}
\caption{Robustness evaluation datasets.}
\label{tab:robust_sets}
\end{table}

\begin{wraptable}{r}{0.4\linewidth}
\centering
\footnotesize
\setlength{\tabcolsep}{4pt}
\renewcommand{\arraystretch}{1.0}
\begin{tabular}{|c|c|c|c|}
\hline
Configuration & ARR & \multicolumn{2}{c|}{RR} \\
(\textsc{Qwen2.5-14B}) &  & Group 1 & Group 2 \\
\hline 
V1 & 3/6 & \textbf{3/3} & 2/3 $\dagger$ 
\\
V6 & \textbf{6/6} & 2/3 & \textbf{2/3} \\
V7 & 4/6 & \textbf{3/3} & 1/3 \\
\hline 
\end{tabular}
\caption{Robustness evaluation ($\dagger$: robust but incorrect translations for that group).}
\label{tab_robust_eval}
\end{wraptable}

The idea behind RR is that when a NL requirement is rephrased without changing its meaning, the resulting LTL formula should not change either. For example, all requirements in Table~\ref{tab:robust_sets} under Group G2, column RR, should be translated to the same LTL formula. Ideally, this LTL formula should be
the correct translation, $\always(\textit{message}\to\eventually\,\textit{delivered})$. But again, we separate robustness from correctness, and qualify incorrect but identical translations as RR robust.

We evaluated \tlguard variants V1, V6 and V7 with  \textsc{Qwen2.5-14B} for ARR and RR. For ARR, we used the requirements from Table~\ref{tab:robust_sets} (ARR). To evaluate RR, we used the two groups of requirements from Table~\ref{tab:robust_sets} (RR with G1 and G2). Results are summarized in Table~\ref{tab_robust_eval}. 
For ARR robustness, V6 and V7 show an improvement over V1, with V6 achieving perfect robustness. 

Interestingly, V1 translates 3 out of 6 requirements to the correct $\always(p\to\eventually q)$ form, and the remaining 3 requirements to the same incorrect form $\always\eventually q$, where $q$ is \textit{delivered}, \textit{shipped}, or \textit{completed}.
V7 translates 4 out of 6 requirements (1, 3, 4, and 5) to the correct $\always(p\to\eventually q)$ form, and the remaining 3 requirements to the same incorrect form $\always\eventually q$. 
V6 translates all 6 requirements to the correct $\always(p\to\eventually q)$ form.

For RR robustness, the vanilla variant V1 already shows strong RR robustness for both groups of requirements, which is not improved 
by neither V6 nor V7.
More specifically,
V1 translates all 3 requirements of G1 to the correct $\always(\textit{request}\to\eventually\,\textit{grant})$ formula, and so does V7,  except that it translates requirement b as $\always(\textit{request}\to\eventually\,\textit{granted})$ (which we consider correct). V6 translates requirements a and c of G1 to the correct formula, but requirement b to the incorrect formula $\always(\textit{request}\to\eventually\always\,\textit{granted})$.
For G2, V1 translates requirements d and f to the same incorrect formula $\always(\eventually\,\textit{delivered})$ and requirement e to the incorrect formula $\always(\next d \to \eventually d)$. So V1 achieves RR robustness 2/3 on G2, even though all its translations are incorrect for this group (hence the $\dagger$ mark).
V6  translates requirements d and f to the correct formula $\always(\textit{message}\to\eventually\,\textit{delivered})$, and requirement e to the incorrect formula $\always(\textit{delivered})$.
V7 translates requirement f to the correct formula, and requirements d and e to the incorrect formulas $\always(\eventually\ \textit{delivered})$ and $\always(\textit{delivered})$, respectively.

\begin{graybox}
\textbf{(RQ2):} \tlguard (esp.  V6) shows an overall robust behavior with respect to the linguistic variants of the requirements, 
but that must be further improved for an industrial-grade use. 
\end{graybox}

\subsubsection{(RQ3) Consistency Checking}

We evaluate the impact of consistency checking to detect potential inconsistencies in the requirements due to either (1) conflicts in the  NL requirements, or (2) the result of translation errors. We ran \tlguard with variants V6 and V7 on the set of input requirements $R1$, $R2$ and $R3$ shown in Table~\ref{tab:req_and_variants} (left).

\begin{table}[tbh]
\centering
\footnotesize
\setlength{\tabcolsep}{4pt}
\renewcommand{\arraystretch}{1.0}

\begin{minipage}[t]{0.48\linewidth}
\centering
\begin{tabular}{|c|p{0.65\linewidth}|}
\hline
\textbf{Req ID} & \textbf{Requirement} \\
\hline
R1 & every request must be granted \\
R2 & requests will not be granted \\
R3 & some request will be made \\
\hline
\end{tabular}
\end{minipage}
\hfill
\begin{minipage}[t]{0.48\linewidth}
\centering
\begin{tabular}{|c|p{0.65\linewidth}|}
\hline
\textbf{Req ID} & \textbf{Requirement} \\
\hline
R2a & requests are not granted \\
R2b & no request shall be granted \\
R2c & requests are never granted \\
R2d & requests shall not be granted \\
R2e & requests will never be granted \\
\hline
\end{tabular}
\end{minipage}

\caption{Set of requirements: (left) original set and (right) variants of requirement R2}
\label{tab:req_and_variants}
\end{table}

$R1$, $R2$ and $R3$ are translated (jointly) to the LTL formula in Equation~\ref{eq:first_formula}.
BLACK finds this formula to be UNSAT, thus detecting a conflict in $R1$, $R2$ and $R3$.\footnote{
$R2$ could be alternatively translated as 
$\always(\textit{request} \to \always(\neg\,\textit{granted}))$. 
Both translations result in inconsistency.
}
Interestingly, varying $R2$ to the seemingly equivalent variants shown in Table~\ref{tab:req_and_variants}~(right) results in different outcomes. Combining $R1$ and $R3$ with $R2a$ results in the correct LTL formula (Equation~\ref{eq:first_formula}). Combining $R1$ and $R3$ with either $R2b$ or $R2d$ results in the translator generating the LTL formula shown in Equation~\ref{eq:second_formula}, which is also UNSAT.
But the inconsistency now stems from the conflict between $\always(\neg\,\textit{request})$ and $\eventually(\textit{request})$, caused by translating $R2b$ or $R2d$ as $\always(\neg\,\textit{request})$ instead of $\always(\neg\,\textit{granted})$. This shows that satisfiability checking helps detect not only inconsistencies in the original NL requirements but also translation errors.
Combining $R1$ and $R3$ 
with either $R2c$ or $R2e$ under V6 yields correct individual LTL formulas for each of the requirements; however, these formulas are incorrectly joined by disjunction instead of conjunction. In contrast, V7 produces the correct joint formula (with conjunction).

{
\setlength{\abovedisplayskip}{-8pt}
\setlength{\belowdisplayskip}{2pt}
\begin{align}
\label{eq:first_formula}
\always(\textit{request} \to \eventually \, \textit{granted})
\;\land\;
\always(\neg\,\textit{granted})
\;\land\;
\eventually(\textit{request})
\\
\label{eq:second_formula}
\always(\textit{request} \to \eventually \, \textit{granted})
\;\land\;
\always(\neg\,\textit{request})
\;\land\;
\eventually(\textit{request})
\end{align}
}

\begin{graybox}
\textbf{(RQ3):} Consistency checking proves to play an essential role in detecting conflicts in requirements, but also in detecting erroneous formalization of requirements. 
\end{graybox}

\section{Conclusions and Future Work}

We introduced \tlguard, a framework for translating informal requirements into LTL specifications using compact language models. We showed that by combining their generative capabilities with lightweight symbolic techniques that provide syntactic and semantic guidance for LTL, the framework achieves strong robustness and translation accuracy, competitive with approaches based on large flagship models. Nevertheless, our results indicate that there remains substantial room for further improving both the robustness and the overall accuracy of the approach.

In the future, we plan a more interactive approach with more accurate explanations of the results. We will investigate methods for resolving translations of ambiguous requirements. Finally, we will consider more expressive formalisms, including LTL with past and Signal Temporal Logic (STL). 
We will finally investigate combining multiple modalities (e.g. requirements consisting both of natural language but also visual examples of the traces satisfying or violating requirements).

\paragraph{Acknowledgments.}
This work was supported by 
(i) the Vienna Science and Technology Fund (WWTF) under Grant 10.47379/ICT23-030; 
(ii) the European Union. Views and opinions expressed are however those of the author(s) only and do not necessarily reflect those of the European Union or the European Health and Digital Executive Agency (HADEA). Neither the European Union nor the granting authority can be held responsible for them. RobustifAI project, ID 101212818; (iii) the U.S. National Science Foundation (NSF) Formal Methods in the Field (FMitF) program under Awards 2319500 and 2525087; and (iv) EU's CYBERGUARD project, Grant No 101190281.

\bibstyle{plain}
\bibliography{refs}   
\end{document}